\documentclass{ws-p8-50x6-00}

\def\Z{{\cal Z}}
\def\L{{\cal L}}
\def\t{\tilde}

\begin{document}
	 
\title{QCD with a $\theta$-vacuum term: a complex system with a 
simple complex action}

\author{V. Azcoiti\footnote{Talk presented by V. Azcoiti}~ and V. Laliena}

\address{Departamento de F\'\i sica Te\'orica, Facultad 
de Ciencias, Universidad de Zaragoza, 50009 Zaragoza, 
Spain\\E-mail: azcoiti@azcoiti.unizar.es, laliena@posta.unizar.es}

\author{A. Galante}

\address{Dipartimento di Fisica dell'Universit\`a di L'Aquila,  
67100 L'Aquila, Italy\\E-mail: galante@lngs.infn.it}


\maketitle

\abstracts{
We reanalyze in the first part of this paper the old question of P and CT 
realization in QCD. The second part is devoted to establish general results 
on the phase structure of this model in the presence of a $\theta$-vacuum 
term.
}

\section{Introduction}
Path integral formulation of Quantum Field Theories associates, in the more 
standard cases, a four-dimensional classical statistical mechanics system 
to a given quantum system. The euclidean partition function shows a 
well-defined Boltzmann weight, the correlation functions are the euclidean 
Green functions and the free energy density is equivalent to the vacuum 
energy density in the hamiltonian approach.

Even if the previous statement is true for almost all physically 
relevant systems, there are some notorious and important exceptions. 
The first relevant exception is lattice formulation of Quantum Chromodynamics 
at finite baryon density. Since the determinant of the Dirac operator at 
finite chemical potential is a complex number, the euclidean action 
of this model is complex. This is the most standard example of a complex 
system with a complex action. The meaning of "complex action" in this case 
is twofold: first, as previously discussed, the action is a complex number 
and second the complex part of the action involves fermionic degrees of 
freedom which are Grassmann variables. It is well known that these two 
features have enormously delayed all attempts to analyze the behaviour of high 
density matter from first principles.

The second relevant exception to the equivalence between a quantum system 
and a classical statistical mechanics system is QCD with a topological term in 
the action. Due to the fact that the local density of topological charge 
picks-up a factor of i under Wick rotation, the euclidean action of this model 
is also complex. This is an example of a complex system with a simple complex 
action. In fact even if the system is complex as opposite to simple, the 
complex part of the action is at least much simpler that the one which 
appears in QCD at finite baryon density since it involves only gauge degrees 
of freedom. It is therefore natural to think that if we have some hope to 
understand QCD at finite baryon density from first principles, we should be 
able to understand previously a simpler case of a complex system with a complex 
action: QCD with a topological term in the action. This was our last but not 
the least motivation to analyze this system.

This paper contains two differentiate parts. In the first one we will give 
a new look to the old Vafa-Witten theorem on the impossibility to break 
spontaneously parity and CT in vector-like theories as QCD\cite{WITTEN} and 
will show how an essential ingredient in the Vafa-Witten demonstration, 
the assumption that the free energy density exists in the presence of any 
external symmetry breaking field, requires the previous assumption that the 
symmetry is realized in the vacuum\cite{PRL}. In other words, the thesis of 
the theorem is in some sense assumed as an hypothesis.

The second part of the paper is devoted to the analysis of the phase structure 
of QCD with a $\theta$-vacuum term in the action. 
We will demonstrate a theorem which stays that if the 
$\theta$-vacuum term is relevant, either QCD has a phase transition at 
$\theta=\pi$ which breaks spontaneously parity or the free energy density 
has some non-analyticity at some $\theta<\pi$.

\vskip 0.6truecm

\section{ New look to the Vafa-Witten theorem.}
Let us start this section by recalling the main ingredients and steps of 
Vafa-Witten theorem\cite{WITTEN} which stays that parity and CT cannot be 
spontaneously broken in vector-like parity conserving theories as QCD.

The theorem is based on the following two main points:

(i) The crucial observation that any arbitrary hermitian local order parameter 
X for parity constructed from Bose fields have to be proportional to an odd 
power of the four indices antisymmetric tensor $\epsilon^{\mu\nu\rho\eta}$,

(ii) The assumption that the free energy density is well defined in the 
presence of a symmetry breaking source $\lambda X$.

The first of these two ingredients implies that any arbitrary order parameter 
X should contain and odd number of time derivatives plus time components 
of the gauge field and should pick-up therefore a factor of $i$ under Wick 
rotation. This observation plus assumption (ii) makes the demonstration of the 
thesis actually very simple. In fact let $\Z$ be the euclidean partition 
function of QCD (or any parity conserving vector-like theory) in the presence 
of a local symmetry breaking source X, 

\begin{equation}
\Z = 
\int dA^{a}_{\mu} d\bar\psi d\psi exp\left( -\int d^{4}x \left( \L(x) + 
i\lambda X(x)\right) \right)
\label{1}
\end{equation}

\noindent
where $\L(x)$ is the standard QCD Lagrangian and $X(x)$ is a real number since 
we have exhibited explicitly the factor of i which arises from Wick rotation. 
As can be seen in (1) the symmetry breaking term is a pure phase factor in 
the integrand of the partition function.

Since the determinant of the Dirac operator is positive definite in a 
vector-like theory as QCD, the presence of a pure phase factor in the 
integrand of the partition function can only decrease the value of $\Z$. 
Therefore the free energy density $f(\lambda)$ defined as

\begin{equation}
\Z = e^{-V f(\lambda)}
\label{2}
\end{equation}

\noindent
where $V$ is the space-time volume, will increase  with $\lambda$. 
The vacuum energy density, which is given by the free energy density, will 
also increase with $\lambda$ and therefore the symmetric vacuum will be 
stable under small perturbations.

Before going on in a deeper analysis let us say that it is actually surprising 
the fundamental role played in the mechanism previously discussed by the 
factor of $i$ picked-up by the order parameter under Wick rotation. In fact 
the standard way to analyze spontaneous symmetry breaking is to add a symmetry 
breaking source to the symmetric action, compute the mean value of the 
order parameter, take the infinite volume limit and then the limit of 
vanishing symmetry breaking source. If at the end of this procedure we 
get a non-vanishing value for the order parameter the symmetry is 
spontaneously broken. But it is also well known that the symmetric action 
contains enough information on the realization of the symmetry in the vacuum. 
Indeed the analysis of the probability distribution function (p.d.f.) of the 
order parameter in the symmetric model has been extensively and successfully 
employed in the investigations on spontaneous symmetry breaking in spin 
systems like the Ising model\cite{BINDER}, in spin-glass 
models in order to analyze 
the complicated structure of equilibrium states not connected by symmetry 
transformations\cite{GIORGIO}, 
and more recently this formalism has also been extended to 
quantum systems with fermionic degrees of freedom\cite{VIC}.

But what is even more surprising is the contradictory result we obtain if 
we apply Vafa-Witten argumentation to the Ising model. The Ising model is 
not a quantum field theory but it verifies the main requirement in Vafa-Witten 
theorem since the integration measure in this model is positive definite.
If we add an imaginary external magnetic field to the hamiltonian of the 
Ising model and apply Vafa-Witten argumentation, we should conclude that the 
$Z_2$ symmetry of this model is not spontaneously broken. But this is 
obviously wrong since it is well known that in the low temperature phase the 
Ising model shows spontaneous magnetization. The solution to this paradox 
lies in the fact that the free energy density in the low temperature phase 
and for an imaginary magnetic field is not defined (it is singular on the 
imaginary axis of the complex magnetic field plane). The Lee-Yang zeroes of 
the partition function\cite{LEE} live on the imaginary axis and approach 
the origin with velocity V, the lattice volume, forbidding to get a 
well defined thermodynamical limit.

We will show in the following that this is not a pathology of the Ising model 
but a general feature of any model with a discrete $Z_2$ symmetry. In other 
words, we will show that to assume the euclidean free energy density 
$f(\lambda)$ is well defined in the presence of any external symmetry 
breaking source requires the previous assumption that the symmetry is 
realized in the vacuum.

To start the proof let us come back to equation (1) which defines the 
euclidean path-integral formula for the partition function. Using the p.d.f. 
of the order parameter X we can write it as

\begin{equation}
\Z(\lambda) = \Z(0) \int d \t{X} P( \t{X} ,V) e^{-i \lambda V \t{X}}
\label{3}
\end{equation}

\noindent
where $V$ in (3) is the space-time volume, $P(\t{X} ,V)$ is the p.d.f. 
of $X$ at a given volume

\begin{equation}
P(\t{X} ,V) = 
{\int dA^{a}_{\mu} d\bar\psi d\psi e^{-\int d^{4}x \L(x)} 
\delta\left(\bar X(A^{a}_{\mu})-\t{X}\right)
\over \int dA^{a}_{\mu} d\bar\psi d\bar\psi e^{-\int d^{4}x \L(x)}}
\label{4}
\end{equation}

\noindent
and

$$
\bar X(A^{a}_{\mu}) = \frac{1}{V}\int d^{4}x X(x)\nonumber
$$

Notice that, since the integration measure in (4) is positive or at least 
semi-positive definite, $P(\t{X},V)$ is a true well normalized p.d.f.

Let us assume that parity is spontaneously broken. In the simplest case in 
which there is no an extra vacuum degeneracy due to spontaneous breakdown 
of some other symmetry, we will have two vacuum states as corresponds to 
a discrete $Z_2$ symmetry. Since $X$ is an intensive operator, the p.d.f. of 
$X$ will be, in the thermodynamical limit, the sum of two $\delta$ 
distributions:

\begin{equation}
\lim_{V\rightarrow\infty}P(\t{X},V) = 
{1\over2}\delta (\t{X}-a)+{1\over2}\delta (\t{X}+a)
\label{5}
\end{equation}

At any finite volume, $P(\t{X},V)$ will be some symmetric function 
($P(\t{X},V)=P(-\t{X},V)$) developing a two peak structure at 
$\t{X}=\pm a$ and approaching (5) in the infinite volume limit.

Due to the symmetry of $P(\t{X},V)$ we can write the partition function as 

\begin{equation}
\Z(\lambda) = 2\Z(0) Re \int^{\infty}_{0}  P(\t{X},V) e^{-i\lambda V\t{X}}
d\t{X}
\label{6}
\end{equation}

\noindent
and if we pick up a factor of $e^{-i\lambda Va}$

\begin{equation}
\Z(\lambda) = 2\Z(0) Re \left(e^{-i\lambda Va}
\int^{\infty}_{0}  P(\t{X},V) e^{-i\lambda V(\t{X}-a)}
d\t{X}\right)
\label{7}
\end{equation}

\noindent
which after simple algebra reads as follows:

\begin{eqnarray}
\label{8}
\Z(\lambda)/(2\Z(0)) =&&
\cos (\lambda Va)\int^{\infty}_{0}  P(\t{X},V) \cos\left(\lambda 
V(\t{X}-a)\right)d\t{X}\nonumber \\
&-& \sin (\lambda Va)\int^{\infty}_{0}  P(\t{X},V) \sin\left(\lambda 
V(\t{X}-a)\right) d\t{X} 
\end{eqnarray}

The relevant zeroes of the partition function in $\lambda$ can be obtained 
as the solutions of the following equation:

\begin{equation}
\cot (\lambda Va)=
{{\int^{\infty}_{0}  P(\t{X},V) \sin\left(\lambda V(\t{X}-a)\right)
d\t{X}}\over
{\int^{\infty}_{0}  P(\t{X},V) \cos\left(\lambda V(\t{X}-a)\right)
d\t{X}}}
\label{9}
\end{equation}

Let us assume for a while that the denominator in (9) is constant at large 
$V$. Since the absolute value of the numerator is bounded by 1, the partition 
function will have an infinite number of zeroes approaching the origin 
($\lambda=0$) with velocity $V$. In such a situation the free energy density 
does not converge in the infinite volume limit.

But this is essentially what happens in the actual case. In fact if we 
consider the integral in (9) 

\begin{equation}
f(\lambda V,V)=
\int^{\infty}_{0}  P(\t{X},V) \cos\left(\lambda V(\bar{X}-a)\right)
d\t{X}
\label{10}
\end{equation}

\noindent
as a function of $\lambda V$ and $V$ it is easy to check that the derivative of 
$f(\lambda V,V)$ respect to $\lambda V$ vanishes in the large volume limit 
due to the fact that $P(\t{X},V)$ develops a $\delta(\t{X}-a)$ 
in the infinite
volume limit. At fixed large volumes $V$, 
$f(\lambda V,V)$ as function of $\lambda V$ is 
an almost constant non-vanishing function (it takes the value of $1/2$ at 
$\lambda V=0$). The previous result 
on the zeroes of the partition function in 
$\lambda$ remains therefore unchanged; it generalizes   
the Lee-Yang theorem on the zeroes of the grand canonical partition 
function of the Ising model in the complex fugacity plane\cite{LEE} 
to any statistical model with a discrete $Z_2$ symmetry.

To illustrate this result with an example, let us take for $P(\t{X},V)$ a 
double gaussian distribution 

\begin{equation}
P(\t{X},V)= {1\over2}\left({V\over\pi}\right)^{1/2}
\left(e^{-V(\t{X}-a)^{2}} + e^{-V(\t{X}+a)^{2}}\right) 
\label{11}
\end{equation}

\noindent
which gives for the partition function 

\begin{equation}
\Z(\lambda) = Z(0) \cos(\lambda Va) e^{-{1\over4}\lambda^{2}V}
\label{12}
\end{equation}

\noindent
and for the mean value of the order parameter 

\begin{equation}
<iX> = {1\over2}\lambda + \tan (\lambda aV) a
\label{13}
\end{equation}

The zeroes structure of the partition function is evident in (12) and 
consequently the mean value of the order parameter (13) is not defined 
in the thermodynamical limit. Notice also that if $a=0$ (symmetric vacuum), 
the free energy density is well defined at any $\lambda$ and then 
Vafa-Witten's argument applies.

\subsection{Conclusions}

We have shown in the previous section that an essential ingredient in the 
Vafa-Witten theorem on the impossibility to break spontaneously parity in 
a parity-conserving vector-like theory as QCD, the existence of the free energy 
density $f(\lambda)$ in the presence of any symmetry breaking external 
source $\lambda X$, does not work if the symmetry is spontaneously broken. 
The requirement that the free energy density is well defined implies the 
previous assumption that the symmetry is realized in the vacuum. This does not 
necessarily implies that Vafa-Witten conjecture is wrong but at least a 
theorem on it is still lacking.

From a speculative point of view let us assume for a while that strong 
interaction shows a weak spontaneous parity breaking. In such a case we 
would expect a non-vanishing vacuum expectation value for the density of 
topological charge at $\theta =0$ and, as we have shown, the theory would 
be ill-defined at $\theta \neq 0$. Actually this issue seems not very likely 
from a phenomenological point of view, but this is not the only scenario in 
which the theory is ill-defined at $\theta \neq 0$. In fact even if the 
symmetry is realized in the vacuum, the partition function could have zeroes 
in $\theta$ approaching the $\theta =0$ point with velocity less than the 
space-time volume V and also in this case the theory would be ill-defined at 
$\theta \neq 0$.

A simple model which shows this feature is the lattice free-fermion theory, 
the partition function of which shows zeroes on the imaginary axis of the 
complex fermion-mass plane approaching the origin with velocity less than 
the lattice volume V. The chiral symmetry is realized in the free fermion 
model and notwithstanding that, the zeroes structure of the partition 
function forbids to define the theory at imaginary values of the fermion mass. 

This speculative mechanism would provide us with a simple explanation for the 
$\theta -vacuum$ or strong CP problem: $\theta$ is zero because otherwise the 
theory is ill-defined. As stated before this is a pure speculation at present 
but we think it is worthwhile to investigate such an issue.

\section{QCD at finite $\theta$.}

The partition function of QCD with a topological term in the action reads
as follows:

\begin{equation}
{\cal Z}\;=\;\int\,[dA_\mu^a]\,[d\psi]\,[d\bar\psi]\,\exp\,
\left\{\,-\int\,d^4x\,
[{\cal L}(x)-\frac{{\mathrm i}\theta}{16\pi^2}X(x)]\,\right\}
\label{QCDpf}
\end{equation}

\noindent
where ${\cal L}(x)$ is the standard QCD Lagrangian and

\begin{equation}
X(x)\;=\;\epsilon_{\mu\nu\rho\sigma}\,{\mathrm Tr}\,
F_{\mu\nu}F_{\rho\sigma}
\label{ffdual}
\end{equation}

\noindent
is, up to a normalization constant, $1/16\pi^2$, the euclidean local density 
of topological charge. The normalization constant has been chosen in such a
way that the topological charge

\begin{equation}
\frac{1}{16\pi^2}\,\int\,d^4x\, X(x)
\label{topcharge}
\end{equation}

\noindent
is an integer.

Assuming that the theory has a non-trivial $\theta$ dependence, we will
show now the following theorem:
{\em ''QCD with a topological term in the action, either breaks parity 
spontaneously at $\theta=\pi$, or has a phase transition at some critical 
$\theta_c$ less than $\pi$''}.

To start with the proof, let us write the partition function 
in a finite space-time 
volume, $V$, as a sum over all topological sectors, labeled by the integer
$n$ that gives the topological charge of the partition functions of
each sector, weighted by the proper topological phase:

\begin{equation}
{\cal Z}_V(\theta)\;=\;\sum_n\,g_V(n)\,{\mathrm e}^{{\mathrm i}\theta n}
\end{equation}

\noindent
where 

\begin{equation}
g_V(n)\;=\;\int_{n}\,[dA_\mu^a]\,[d\psi]\,[d\bar\psi]\,\exp[-\int\,d^4x\,
{\cal L}(x)]
\end{equation}

\noindent
is the standard partition function computed over the gauge sector with
topological charge equal to $n$. The function $g_V(n)$ is, up to a 
normalization factor, $\sum_n g_V(n)$, the probability, $p_V(n)$, of
the topological sector $n$ at $\theta=0$:

\begin{equation}
p_V(n)\;=\;\frac{g_V(n)}{\sum_n g_V(n)} \, .
\end{equation}

If we define the mean topological charge density

\begin{equation}
x_n\;=\;\frac{n}{V}
\end{equation}

\noindent
we can write the previous partition function as

\begin{equation}
{\cal Z}_V(\theta)\;=\;\sum_{x_n}\,h_V(x_n)\,
{\mathrm e}^{{\mathrm i}\theta V x_n} \, ,
\label{pfdiscrete}
\end{equation}

\noindent
where $h_V(x_n)=g_V(n)$ and the step $\Delta x_n$ in the
topological charge density is $1/V$.

Let the new $h_V(x)$ be a continuous interpolation of
$h_V(x_n)$ and let us define a new function of $\theta$ in the
following way:

\begin{equation}
{\cal Z}_{c,V}(\theta)\;=\;\int\,dx\,h_V(x)\,
{\mathrm e}^{{\mathrm i}\theta Vx} \, .
\label{pfcontinuum}
\end{equation}

Summing up the ''pseudo-partition'' functions 
${\cal Z}_{c,V}(\theta+2\pi m)$ for all integers $m$ we get

\begin{equation}
\sum_m\,{\cal Z}_{c,V}(\theta+2\pi m)\;=\;
\int\,dx\,h_V(x)\,{\mathrm e}^{{\mathrm i}\theta Vx}\,
\sum_m\,{\mathrm e}^{{\mathrm i}2\pi mVx}\, .
\end{equation}

\noindent
Using now the following representation of the periodic
delta function:

\begin{equation}
\sum_m\,{\mathrm e}^{{\mathrm i}2\pi mVx}\;=\;\frac{1}{V}\,
\sum_m\,\delta(x-\frac{m}{V}) \, ,
\end{equation}

\noindent
we get the following identity:

\begin{equation}
{\cal Z}_V(\theta)\;=\;V\,
\sum_m\,{\cal Z}_{c,V}(\theta+2\pi m) \, ,
\label{vacios}
\end{equation}

\noindent
which relates our QCD partition function ${\cal Z}_V(\theta)$
to the ''pseudo-partition'' functions ${\cal Z}_{c,V}(\theta+2\pi m)$.
The usefulness of this identity will become evident in a while.
Before, let us come back to expression (\ref{pfdiscrete}). 
Under a parity
transformation, the density of topological charge, $x_n$, changes
sign, whereas $h_V(x_n)$ is an even function of $x_n$ since at
$\theta=0$ the QCD action is parity invariant. Taking also into
account that the full topological charge $Vx_n$ is an integer, 
we conclude that at $\theta=\pi$ each term entering the sum
of Eq. \ref{pfdiscrete} is parity invariant. 
Therefore, at $\theta=\pi$ the 
theory recovers the symmetry under parity transformations, and
if the symmetry is also realized in the vacuum we must obtain
a vanishing value for the expectation value of the topological 
charge density: $\langle x\rangle=0$.

Using Eq. (\ref{vacios}) we can write for the mean value of the 
density of topological charge:

\begin{equation}
\langle x\rangle\;=\;\sum_m\,\langle x\rangle_{c,m}\,
\left(\frac{{\cal Z}_{c,V}(\theta+2\pi m)}
{\sum_n{\cal Z}_{c,V}(\theta+2\pi n)}\right) \, ,
\label{avtopcharge}
\end{equation}

\noindent
with

\begin{equation}
\langle x\rangle_{c,m}\;=\;
\frac{\int\,dx\,x\,h_V(x)\,
{\mathrm e}^{{\mathrm i}(\theta+2\pi m)Vx}}
{\int\,dx\,h_V(x)\,
{\mathrm e}^{{\mathrm i}(\theta+2\pi m)Vx}} \, ,
\label{contitopch}
\end{equation}

\noindent
Since $x$ in (\ref{contitopch}) is a continuous variable
(i.e. there is no symmetry forcing the numerator to be zero), 
we expect a non-vanishing value for $\langle x\rangle_{c,m}$
at $\theta=\pi$. However, it is simple to see that parity
symmetry is realized at $\theta=\pi$ in a finite volume.
In fact, the contribution of the $m=0$ sector compensates
the contribution of $m=-1$ in (\ref{avtopcharge}), 
the contributions $m=1$ 
cancels that of $m=2$, and so on. Therefore, at $\theta=\pi$
different sectors compensate to give $\langle x\rangle=0$.

Equation (\ref{vacios}) is valid for any value of $\theta$. 
At $\theta=0$
it is simple to verify that in the infinite volume limit the 
sector $m=0$ gives all the contribution to the partition function.
This is a simple feature which follows from the fact that at
$\theta=0$ and in the infinite volume limit the exact solution
for the free energy density is given by the saddle point solution, which
is also the solution for the $m=0$ sector. If this sector dominates
for any value of $\theta$, we will get a first order phase
transition at $\theta=\pi$, with a non-vanishing value of the
topological charge density (remember the discussion following
Eq. (\ref{contitopch})), and the theory will undergo spontaneous parity 
breaking. Otherwise, there will be some critical value, $\theta_c$,
at which other sectors start to give a contribution to the partition
function, and in such a case we will get a phase transition at this
$\theta_c$. On general grounds, we can say nothing about the number 
and order of phase transitions expected in this case.

To see how this mechanism works in practical cases and to get 
intuition of what we can expect in physical systems, let us analyze
in the following two simple examples: the Ising model within an
imaginary external magnetic field and a gaussian model.

\subsection{The one-dimensional Ising model in an imaginary external 
magnetic field}

In the previous demonstration we have not made use of any specific
property of QCD, except the quantization of the topological charge.
Therefore, the result applies to any model with a quantized topological 
charge which appears as an imaginary contribution to the euclidean action.
A simple example of that is the one-dimensional Ising model in an 
imaginary external magnetic field. The hamiltonian of this model can be 
written as

\begin{equation}
H_N\;=\;-\,J\,\sum_{i=1}^N\,S_i\,S_{i+1}\:-\:{\mathrm i}\,
\frac{\theta k_B T}{2}\,\sum_{i=1}^N\,S_i \, ,
\end{equation}

\noindent
where $J$ is the coupling constant between nearest neighbours,
$k_B$ is the Boltzmann constant, $T$ the physical temperature, $N$ the 
number of spins, and we assume periodic boundary conditions.

The partition function is given by

\begin{equation}
{\cal Z}_N\;=\;\sum_{\{S\}}\,{\mathrm e}^{F\sum_i S_iS_{i+1}\,+\,
{\mathrm i}\theta\frac{1}{2}\sum_i S_i}
\label{pfising}
\end{equation}

\noindent
where $F=J/k_B T$ and the sum is over all spin configurations.

For an even number of spins, the quantity $1/2\,\sum_i S_i$ which
appears in the imaginary part of the hamiltonian is an integer
taking values between $-N/2$ and $N/2$, and therefore it can
be seen as a quantized ''topological'' charge. Furthermore, the 
theory has a $Z_2$ symmetry at $\theta=0$ and $\theta=\pi$ which,
in the sense of this work, is the analogous of parity in QCD.

The transfer matrix technique allows to compute exactly the 
partition function defined in Eq. (\ref{pfising}). The final result is

\begin{equation}
{\cal Z}_N\;=\;\lambda_+^N\:+\:\lambda_-^N \, ,
\end{equation}

\noindent
where $\lambda_\pm$ are the two eigenvalues of the transfer matrix,
which are given by the following equation:

\begin{equation}
\lambda_\pm(\theta)\;=\;{\mathrm e}^F\,\cos\frac{\theta}{2}\:\pm\:
\left(-{\mathrm e}^{2F}\,\sin^2\frac{\theta}{2}\,+\,{\mathrm e}^{-2F}
\right)^{1/2} \, .
\label{eigenvalues}
\end{equation}

\noindent
We see that the external field $\theta$ is an angle taking values
between $-\pi$ and $\pi$, as expected.

Let us discuss first the simpler infinite temperature case. 
At $T=\infty$, $\lambda_-=0$ and $\lambda_+=2\cos(\theta/2)$. The
partition function has the simple form:

\begin{equation}
{\cal Z}_N\;=\;2^N\,[\cos(\theta/2)]^N \, ,
\end{equation}

\noindent
from which it follows that the free energy density is

\begin{equation}
{f}\;=\;\frac{1}{N}\,\log{\cal Z}_N\;=\;\log 2 \:+\:
\log\cos\frac{\theta}{2} \, .
\end{equation}

\noindent
For the mean magnetization we get

\begin{equation}
\langle m\rangle\;=\;
\left\langle\frac{1}{N}\sum_i S_i\right\rangle
\;=\;{\mathrm i}\,\tan\frac{\theta}{2} \, .
\end{equation}

\noindent
This last expression shows that a first order phase 
transition takes place at $\theta=\pm\pi$, with
divergent spontaneous magnetization. The origin of
this phase transition is clarified by noticing that
the sector $m=0$ in Eqs. (\ref{vacios}) and 
(\ref{avtopcharge})
dominates the thermodynamical limit for any $\theta$.

The finite temperature case is somehow different. The 
eigenvalues of the transfer matrix, given by 
Eq. (\ref{eigenvalues}), are real and positive if

\begin{equation}
\sin^2\frac{\theta}{2}\;\leq\;{\mathrm e}^{-4F} \, .
\end{equation}

\noindent
In this case the free energy density has a well defined
thermodynamical limit, and the mean magnetization is
given by

\begin{equation}
\langle m\rangle\;=\;{\mathrm i}\,
\frac{\sin(\theta/2)}
{[{\mathrm e}^{-4F}\,-\,\sin^2(\theta/2)]^{1/2}} \, ,
\label{magnet}
\end{equation}

\noindent
which shows a first order phase transition with a
divergent mean magnetization at

\begin{equation}
\theta_c^\pm\;=\;\pm\,2\,\arcsin {\mathrm e}^{-2F} \, .
\end{equation}

For $\theta_c^+<\theta<\pi$ or 
$-\pi<\theta<\theta_c^-$, the two eigenvalues
of the transfer matrix are complex conjugate numbers.
The partition function ${\cal Z}_N$ is real but not
positive definite, and oscillates in sign with $N$,
making it impossible to define a thermodynamical 
limit. For instance, the mean magnetization does not
converge as $N\rightarrow\infty$. Thus, the theory is 
ill-defined in these intervals of $\theta$.
For $\theta_c^-<\theta<\theta_c^+$, the sum in the 
right-hand side of Eqs. (\ref{vacios}) and 
(\ref{avtopcharge})
is dominated by the term with $m=0$, and the mean
magnetization (\ref{magnet}) is the analytic continuation of 
the saddle point solution obtained with real magnetic
field.

\subsection{ Gaussian distribution.}

The second illustrative example we want to discuss here 
are models with a density of topological charge at $\theta=0$
distributed according to a gaussian distribution.
Thus, let us assume that
the function $h_V(x_n)$ which enters Eq. (\ref{pfdiscrete}) 
has the form

\begin{equation}
h_V(x_n)\;=\;{\mathrm e}^{-V a x_n^2} \, ,
\end{equation}

\noindent
where $a$ is a parameter related to the width of the 
distribution. This form of $h_V(x_n)$ is a natural assumption
from a physical point of view as a first approximation to the
actual distribution of nearly any model. In fact, outside
second order phase transitions, the probability distribution 
function of intensive operators as the density of topological
charge is expected to be gaussian in the vicinity of its maximum.
Of course, deviations from the gaussian behaviour far from the
maximum can induce important changes in the $\theta$-dependence
of the theory in the large $\theta$ regime (the previous
example illustrates very well this point). However, the gaussian
distribution provides us with a simple model that can be
analytically solved and gives useful insights on the general
problem that we are addressing.

The partition function of the model is then

\begin{equation}
{\cal Z}_V(\theta)\;=\;\sum_{x_n}\,{\mathrm e}^{-Vax_n^2}
{\mathrm e}^{{\mathrm i}\theta V x_n} \, .
\end{equation}

The pseudo-partition function ${\cal Z}_{c,V}(\theta+2\pi m)$
entering Eq. (\ref{vacios}) can be analytically computed, and reads:

\begin{equation}
{\cal Z}_{c,V}(\theta+2\pi m)\;=\;\int\,dx\,{\mathrm e}^{-Vax^2}
{\mathrm e}^{{\mathrm i}(\theta+2\pi m)Vx} \;=\;
\left(\frac{\pi}{aV}\right)^{1/2}\,
{\mathrm e}^{-\frac{1}{4a}(\theta+2\pi m)^2 V} \, .
\end{equation}

Using Eq. (\ref{vacios}) we can write

\begin{equation}
{\cal Z}_V(\theta)\;=\;\left(\frac{\pi V}{a}\right)^{1/2}\,
\sum_m\,{\mathrm e}^{-\frac{1}{4a}(\theta+2\pi m)^2 V} \, ,
\end{equation} 

\noindent
and, if $|\theta|<\pi$, it is simple to verify that the free 
energy density ${f}(\theta)$ is given by

\begin{equation}
{f}(\theta)\;=\;\lim_{V\rightarrow\infty}\,
\frac{1}{V}\,\log\,{\cal Z}_V(\theta)\;=\;-\frac{1}{4a}\,
\theta^2 \, .
\label{fegaussian}
\end{equation}

The $m=0$ sector dominates for every $\theta$ between $-\pi$ 
and $\pi$. The vacuum expectation value of the density of
topological charge, $\langle x\rangle$, is

\begin{equation}
\langle x\rangle\;=\;{\mathrm i}\frac{\theta}{2a} \, ,
\label{tcgaussian}
\end{equation}

\noindent
and the model breaks spontaneously parity at $\theta=\pi$.

This simple model which, as we have seen, illustrates very
well the theorem demonstrated in the first part of this
section, has also further relevance. In fact, using a duality
relation of gauge theories with a large number of colours
and string theory on a certain space-time manifold, Witten
has recently studied the $\theta$ dependence of pure gauge
theories in four dimensions\cite{WITTEN2} and has found Eqs.
(\ref{fegaussian}) and (\ref{tcgaussian}). 
Combining both results, we conclude 
that the probability distribution function of the topological
charge density in pure $SU(N)$ gauge theory is gaussian in
the large $N$ limit.
It is interesting to note that also another analytically solvable
model (although not physically relevant)
has the same $\theta$ dependence: the quantum rotor 
in one dimension\cite{ROTOR}.

\subsection{Conclusions.}

We have demonstrated in the previous sections a theorem
which states that if the $\theta$ dependence of QCD is
non-trivial, either the theory has a first order phase 
transition at $\theta=\pi$ which breaks parity spontaneously,
or the model shows a phase transition at some $\theta_c$ less 
than $\pi$. In the last case, we are not able to establish the
order of the phase transition. Furthermore, the proof makes no
use of any specific property of QCD, apart from the quantization 
of the topological charge. This implies that the result applies
to any vector-like model with a quantized charge that appears
as an imaginary contribution to the euclidean action.

To illustrate this result in practical cases and to get intuition
on what we can expect in physical models, we have analyzed two
simple examples: the one-dimensional Ising model in an imaginary
external magnetic field, and a model in which the probability 
distribution function of the density of topological charge at 
$\theta=0$ is assumed to be gaussian.

In the first case, we have found a first order phase transition
with divergent spontaneous magnetization, at a critical imaginary
magnetic field that depends on the temperature. At infinite 
temperature the transition is located at $\theta_c=\pi$ and the 
system breaks spontaneously its $Z_2$ symmetry at this $\theta$
value. At finite temperature the transition appears at 
$\theta_c<\pi$, and the theory is ill-defined for larger values
of $\theta$ (modulo $2\pi$).

The model with gaussian distribution can be analyzed simply, and
we have found a first order phase transition at $\theta_c=\pi$,
which breaks spontaneously parity. The free energy density
and the density of topological charge, which are quadratic and 
linear functions of $\theta$ respectively, agree exactly with the
$\theta$-dependence of $SU(N)$ gauge theories in the large $N$ limit
found recently by Witten. This implies that the topological
charge density of this theory has a gaussian distribution at
$\theta=0$.

\end{document}